\documentclass[preprint,12pt]{elsarticle}
\usepackage{graphicx}
\usepackage{multirow}
\usepackage{caption}
\usepackage{graphicx}
\usepackage{subfigure}
\usepackage{amssymb}
\usepackage{amsfonts,amsmath,amssymb,amscd}
\usepackage{amsthm}
\usepackage{txfonts}
\usepackage{amsthm}
\theoremstyle{plain}

\theoremstyle{definition}
\newtheorem{definition}{Definition}[section]

\newtheorem{remark}{Remark}
\theoremstyle{remark}
\usepackage[english]{babel}
\usepackage{amssymb,amstext,amsmath}
\usepackage{amsfonts,amsmath,amssymb,amscd}
\usepackage{amsfonts}
\usepackage{mathrsfs}
\usepackage{lineno}
\journal{ELSEVIER}

\begin{document}

\begin{frontmatter}

\title{Portfolio selection models based on interval-valued conditional value at risk (ICVaR) and empirical analysis}

 \author[label1]{Jinping Zhang\corref{cor1}}
 \address[label1]{School of Mathematics and Physics, North China Electric Power University,
Beijing, 102206, P. R. China}
\cortext[cor1]{Corresponding author}
\ead{zhangjinping@ncepu.edu.cn}
\author[label1]{Keming Zhang}
\ead{k.m.zhang@ncepu.edu.cn}

\begin{abstract}
Risk management is very important for individual investors or companies. There are many ways to measure the risk of investment.  Prices of risky assets vary rapidly and randomly due to the complexity of finance market. Random interval is a good tool to describe uncertainty including both randomness and imprecision. Considering the uncertainty of financial market, we employ random intervals to describe the return of a risk asset and define an interval-valued risk measurement, which considers the tail risk. It is called the interval-valued conditional Value- at-Risk (ICVaR, for short).  Such a risk measure satisfies the sub-additivity. Under the new risk measure ICVaR, as a manner similar to the classical portfolio model of Markowitz, two optimal interval-valued portfolio selection models are built. Based on the real data from mainland Chinese stock market, the case study shows that our models are interpretable and consistent with the practical scenarios.

\end{abstract}


\end{frontmatter}

\section{Introduction}
\label{author_sec:1}A portfolio is a collection of stocks, bonds, and financial derivatives held by
investors or financial institutions to spread risk. Since H. Markowitz published the pioneering paper on mean-variance model \cite{Mar} in 1952, there have been a large number of papers on the applications and extension of the classical models. It is known that one major drawback of Markowitz's model is that the solution are sensitive to the underlying parameters, i.e. mean and covariance matrix of assets. In order to overcome the sensitivity, some robust portfolio models are proposed, for example, \cite{Blan} and references therein.   

Risk measurement is very important for investors or financial analysts and must be considered in portfolio selection models. Besides variance, there are many other ways to measure the risk of assets. For example, Konnor and Yamazaki \cite{Kon} (1991) studied the optimal portfolio problem based on  absolute deviation or semi-variance, which is easy to calculate than variance of portfolio.  Value at risk (VaR, for short) is a financial risk measurement that has been
developed and widely used (See e.g. \cite{Cas,Cos,Ram}. Hamel et al. \cite{Ham} (2013) extended real-valued VaR to set-valued case. VaR is convenient, easy to calculate. But VaR has no nice mathematical properties such as convexity, sub-additivity, which are crucial to solve the optimal problem. In order to overcome the drawback of VaR, one can consider the conditional value at risk (CVaR, for short), the coherent risk measurement with nice mathematical properties \cite{Man07,Rock1999}.   Because of the asymmetry of information and the incompleteness of financial market,
the price of risky assets shows a more complex form than a scalar variable. In order to describe the
complexity, randomness and inaccuracy of the financial market, the tool of interval-valued or fuzzy variable is a good choice. For example, Ida \cite{Ida2003} (2003) studied portfolio selection with interval coefficients. \cite{Ida2004} (2004) considered portfolio problem with both interval and fuzzy coefficients. Giove et al. \cite{Gio} (2006) studied the interval portfolio model based on regret function. Zhang and Li \cite{Zha} (2009) considered the portfolio selection models based on interval-valued semi-variance. Hu et al. \cite{Hu} (2020) proposed optimal interval-valued portfolio model based on possibility measure. Yan et al. \cite{Yan} (2017) built non-linear portfolio selection model with interval-valued downside semi-variance constraints. Mohagheghi et al. \cite{Mo} (2017) employed interval type-2 fuzzy optimization approach to study the problem of  research and development project evaluation and project portfolio selection.  Also, there are some references deal with the portfolio selection without using the usual risk measure such as variance, semi-variance, VaR and CVaR. For example, Mansini and Speramza \cite{Man99} (1999) built the portfolio model with minimum transaction lots, which is independent of the risk function.  Under interval framework, one can also consider interval-valued return, interval-valued VaR (IVaR, for short), interval-valued CVaR ( See e.g. \cite{Liz}, 2017). Based on the risk measurement ICVaR, it is natural to study interval-valued portfolio selection. Comparing with existing references of interval-valued portfolio selection models, we shall use ICVaR to depict risk instead of others such as semi-absolute variance or semi-variance etc.

In this paper, we collect the closing price, the highest price and the lowest price of stocks on each trade day. Then the range of return of risky asset can be  obtained as an interval-valued random variable. We use the interval-valued CVaR to describe the risk. Based on the ICVaR, two interval-valued optimal portfolio selection models are established. One is the optimal portfolio by maximizing the return for a given level of risk and the other is the optimal portfolio by minimizing the risk for a given level of return. The interval-valued case is a natural extension of the classic portfolio model, which can comprehensively reflect the complexity of financial market and risk appetite of investors.

The rest of the paper is organized as follows: Section 2 is the preliminaries of interval-valued variables and interval-valued programming. Section 3 contributes to the interval-valued  portfolio selection models based on ICVaR.  Section 4 is devoted to the application of our models in real financial market. Section 5 is the concluding remarks.
In this section, we list the knowledge of interval-valued variables and interval-valued programming needed later.
\label{author:sec:2}
\subsection{Operations and order of intervals}
Let $\mathbb{R}$ be the set of all real numbers, $A, B$ real closed intervals. Let
$A=[a^{L}, a^{U}]=\{a\in\mathbb{R}: a^{L}\leq
a\leq {a^{U}}\}$,  $B=[\  b^{L}, \  b^{U}]=\{b\in\mathbb{R}: \  b^{L}\leq
b\leq {\  b^{U}}\}$. If $a^{L}=a^{U}=a$,
the interval $A=[a, a]$ degenerates to a real number.
$$
A+B:=[a^{L}+\  b^{L}, a^{U}+\  b^{U}],
$$

$$
\lambda \cdot A=\lambda[a^{L}, a^{U}]:=\left\{
\begin{array}{cc}
	[\lambda a^{L}, \lambda a^{U}], \ \ & \ \ {\rm if} \ \ \lambda\geq0,\\
	
	[\lambda a^{U}, \lambda a^{L}], \ \ & \ \ {\rm if} \ \
	\lambda<0.
\end{array}
\right.
$$
Let $m(A):=\frac{1}{2}(a^{L}+a^{U}), \  w(A):=\frac{1}{2}(a^{U}-a^{L})$, which are called the mean and width of the interval $A$, respectively. Then $A$ also can be determined by $m(A)$ and $w(A)$.

Ranking intervals is a crucial step to study interval-valued optimization problems. There are infinitely many ways to rank intervals \cite{Wang}. The popular and natural way is using the endpoints or their functions such as mean and width. In this paper, we use the rule ``mean-first, left-second" to rank intervals for calculating the IVaR and ICVaR. E.g. for two intervals $A=[a^{L},a^{U}]$ and $B=[\  b^{L},\  b^{U}]$, define $A\prec B$ if $m(A)<m(B)$. If $m(A)=m(B)$ and $a^{L}<\  b^{L}$, then  $A\prec B$. $A=B$ if and only if $m(A)=m(B)$, $a^{L}=\  b^{L}$. Multiple intervals can be ranked well according to this rule.

\subsection{Interval-valued optimization}
We shall employ the method introduced in \cite{Sen} (2001) to solve interval-valued optimization portfolio. In \cite{Sen}, at first the authors defined an index $\Gamma$ to describe the acceptability for ranking two intervals. Let $\circledless $ be a partial order of intervals. For any closed intervals $A$ and $B$, define
\begin{eqnarray}
	\Gamma(A\circledless B)=\frac{m(B)-m(A)}{w(B)+w(A),}
\end{eqnarray}
where $w(B)+w(A)\neq0$. The value of function $\Gamma(A\circledless
B)$, denoted by $\gamma$, may be interpreted as the grade of acceptability of `the interval $A$ is less than the interval
$B$'.

The grade of acceptability of $A\circledless B$ may be classified and
interpreted further on the basis of comparative position of mean
and width of interval $B$ with respect to those of interval $A$
as follows:
\begin{eqnarray}
	\Gamma(A\circledless B)=\left\{
	\begin{array}{ll}
		\leq0, \ \ & \ \ {\rm if} \ \ m(B)\leq m(A),\\
		>0, <1, \ \ & \ \ {\rm if} \ \ m(A)<m(B) \ \ and \ \ a^{U} > \  b^{L}\\
		\geq1, \ \ & \ \ {\rm if} \ \ m(A)<m(B) \ \ and \ \ a^{U}\leq\  b^{L}.
	\end{array}
	\right.
\end{eqnarray}
If $\Gamma (A\circledless B)\leq 0$, then the premise `$A$ is
less than $B$'
is not accepted. If $0<\Gamma (A\circledless B)<1$, then the
interpreter accepts the premise `$A$ is less than $B$'
with different grades of satisfaction ranging from zero to one
(excluding zero and one). If $\Gamma (A\circledless B)\geq1$, the
interpreter is absolutely satisfied with the premise `$A\circledless B$'.

\subsection{A satisfactory crisp equivalent system of $Ax\leq B$}
Let $A=[a^{L}, a^{U}], B=[\  b^{L},
\  b^{U}]$ and $x$ be a nonnegative singleton variable. In order
to solve linear programming problems with interval coefficients,
Sengupta et al. \cite{Sen} proposed a satisfactory crisp
equivalent form for the interval inequality relations $Ax\leq B$ or
$Ax\geq B.$

The
satisfactory crisp equivalent form of $Ax\leq B$ is given  as follows (\cite{Sen}):
\begin{equation}\label{leq}
	Ax\leq B \ iff \  \left\{
	\begin{array}{l}
		a^{U}x\leq \  b^{U},\\
		\Gamma(B\circledless Ax)\leq \gamma \in[0,1],
	\end{array}
	\right.
\end{equation}
where $\gamma$ may be interpreted as an optimistic threshold given
and fixed by the decision maker.
Similarly, for $Ax\geq B$, the satisfactory crisp equivalent form
is:
\begin{equation}\label{geq}
	Ax\geq B \ iff \  \left\{
	\begin{array}{l}
		a^{L}x\geq \  b^{L},\\
		\Gamma(Ax\circledless B)\leq \gamma \in[0,1].
	\end{array}
	\right.
\end{equation}
Now we consider  the maximizing interval-valued linear programming problem:
\begin{equation}\label{max}
	\begin{array}{ll}
		Maximize_{\circledless} \ \ & Z={\sum\limits_{i=1}^n}
		[c^{L}_{i},
		c^{U}_{i}]x_{i},\\
		subject \ to \ \ &
		{\sum\limits_{i=1}^n}[a^{L}_{ij},a^{U}_{ij}]
		x_{i}\leq [b^{L}_{j}, b^{U}_{j}],\\
		\ \              &  j=1,2,\cdots, k\\
		\ \              & x_{i} \geq 0,  i=1,\cdots,n.
	\end{array}
\end{equation}

According to \cite{Sen} ,the principle of $\gamma$-index
indicates that for the maximization (minimization) problem, an
interval with a higher midpoint is superior (inferior) to an
interval with a lower mid-value. Therefore, in order to obtain the
max (min) of the interval-valued objective function, considering the
midpoint is our primary concern. We reduce the interval objective
function its central value and use conventional LP (Linear
Programming) techniques for its solution. We also consider the width
but as a secondary attribute, only to confirm whether it is within
the acceptable limit of the decision maker. If it is not, one has to
reduce the extent of width (uncertainty) according to his
satisfaction and thus to obtain a less wide interval from the
non-dominated alternatives accordingly. So by the satisfactory crisp equivalent form \eqref{leq}, the following LP problem
\eqref{maxreal} is the necessary equivalent form of the model \eqref{max}:
\begin{eqnarray}\label{maxreal}
	\begin{array}{ll}
		Maximize \ \ & m(Z)=\frac{1}{2}
		{\sum\limits_{i=1}^n}(c^{L}_{i}+c^{U}_{i})x_{i},\\
		subject \ to \ \ & {\sum\limits_{i=1}^n} a^{U}_{ij}
		x_{i}\leq  b^{U}_{j}, \\
		\ \  \ \ &
		{\sum\limits_{i=1}^n}[(a^{L}_{ij}
		+{a^{U}}_{ij})-\gamma (a^{U}_{ij}-
		a^{L}_{ij})]x_{i} \leq b^{L}_{j}+  b^{U}_{j} + \gamma (b^{U}_{j}-
		b^{L}_{j}) ,\\
		\ & \ j=1,\cdots,k\\
		\ \ \ \ &  x_{i} \geq 0,\ for \ i=1,\cdots,n.
	\end{array}
\end{eqnarray}

Similarly,  the minimizing interval linear programming can be constructed as follows:
\begin{equation}\label{min}
	\begin{array}{ll}
		Minimize_{\circledless} \ \ & Z={\sum\limits_{i=1}^n}
		[c^{L}_{i},
		c^{U}_{i}]x_{i},\\
		subject \ to \ \ &
		{\sum\limits_{i=1}^n}[a^{L}_{ij},a^{U}_{ij}]
		x_{i}\geq [b^{L}_{j},b^{U}_{j}], \ for\ j=1,\cdots,k \\
		& x_{i} \geq 0, i=1,\cdots, n.
	\end{array}
\end{equation}
According to \eqref{geq}, the interval-valued programming problem \eqref{min} can be transformed to the following real valued linear programming model \eqref{minreal}:
\begin{eqnarray}\label{minreal}
	\begin{array}{ll}
		Minimize \ \ & m(Z)=\frac{1}{2}
		{\sum\limits_{i=1}^n}(c^{L}_{i}+c^{U}_{i})x_{i},\\
		subject \ to \ \ & {\sum\limits_{i=1}^n} a^{L}_{ij}
		x_{i}\geq  b^{L}_{j}, \\
		\ \  \ \ &
		{\sum\limits_{i=1}^n}[(a^{L}_{ij}
		+{a^{U}}_{ij})+\gamma (a^{U}_{ij}-
		a^{L}_{ij})]x_{i} \geq b^{L}_{j}+  b^{U}_{j} - \gamma (b^{U}_{j}-
		b^{L}_{j}) ,\\
		\ & \ j=1,\cdots,k\\
		\ \ \ \ &  x_{i} \geq 0,\ for \ i=1,\cdots,n.
	\end{array}
\end{eqnarray}

\noindent {\bf Note:}
In above, as you can see,  there are three symbols `$\prec$', `$\circledless$' and `$\leq$' for interval-comparison. `$\prec$' is the `mean-first, left-second' order of interval for calculating the value of IVaR and ICVaR. `$\circledless$' and `$\leq$' are used together to transfer the interval-valued programming models \eqref{max} and \eqref{min} into real valued linear programming problems \eqref{maxreal} and \eqref{minreal} respectively.  
\section{Portfolio selection based on ICVaR}
\label{author_sec:3}
We consider the portfolio problem with $n$ risky assets. Due to the volatility and complexity of financial market, the profit is uncertain. There are several tools to describe uncertainty, for example, random variable, fuzzy variable, set-valued variable etc. Here the return is considered as a random interval, which includes not only randomness but also imprecision.
\subsection{Interval-valued risk measure}
\subsubsection{Interval-valued random variable}
Let $(\Omega, \mathcal F, P)$ be a complete probability space. $\mathbb R$ be the real space equipped with the Borel sigma-algebra $\mathcal B(\mathbb R)$. $K(\mathbb R)$ ($ K_{c}(\mathbb R)$ )denotes the family of all nonempty closed (nonempty, closed and convex, respectively) subsets of $\mathbb R$ . The mapping $V:\Omega\rightarrow K(\mathbb R)$ is called {\em measurable set-valued mapping} or {\em random set} if for any element $B\in \mathcal B (\mathbb R)$, $\{\omega\in\Omega: V(\omega)\cap B\ne\emptyset\}\in \mathcal F$.

$V$ is called an {\em interval-valued random variable} or {\em random interval} if $V$ is measurable and takes value in $K_{c}(\mathbb R)$.

For any $X\in K_{c}(\mathbb R)$, the {\em distribution function} of $V$ is defined by
$$
F(X):=P(\omega\in\Omega: V(\omega)\prec X),
$$
where $\prec$ is the partial order according to the rule `mean-first, left-second'. $V$ is called {\em integrably bounded} if $\int_{\Omega}\sup_{x\in V(\omega)}|x|dP<\infty$. The real valued random variable $f$ is called an {\em integrable selection} of $V$ if $f(\omega)\in V(\omega) \ a.s.$ and $\int_{\Omega}fdP<\infty$.

For an integrably bounded random set $V$,
$$
S^{1}_{V}:=cl\{f\in L^1(\Omega, \mathbb R): f(\omega)\in V(\omega) \ a.s.\},
$$
where $L^1(\Omega, \mathbb R)$ denotes the family of all integrable real valued random variables, the closure $cl$ is taken in $L^1(\Omega, \mathbb R)$. For two random sets $V$ and $U$, it is well known that $V=U \ a.s.$ if and only if $S^{1}_{V}=S^{1}_{U}$ (\cite{Hia}).

The {\em expectation} of $V$  is defined by
$$
E(V)=\{E(f), f\in S^1_{V}\},
$$
which is closed in $\mathbb R$.


By Lemma 3.1 of \cite{zhang09}, a random set $V$ is an interval-valued random variable if and only if $V=[f, g]$, where both $f$ and $g$ are real valued random variables.

For an integrably bounded interval-valued random variable $V=[f, g]$, obviously it holds that $E(V)=[E(f), E(g)]$

\subsubsection{IVaR and ICVaR}
As a manner similar to the usual  Value-at-Risk and conditional Value-at-Risk, we define the interval-valued Value-at-Risk (IVaR) and interval-valued conditional Value-at-Risk (ICVaR) as following:

\begin{definition}
	Let $(\Omega, \mathcal F, P)$ be a complete probability space. $R$ is the random interval-valued return of a risky asset. Given confidence level $1-\alpha \  (0<\alpha<1)$, the Value-at-Risk is defined by
	$$
	IVaR:=-\inf\{X\in K_{c}(\mathbb R): P(R\prec X)=1-\alpha\},
	$$
	where $\inf$ is taken according to the order $\prec$.
\end{definition}

\begin{definition}
	Let $(\Omega, \mathcal F, P)$ be a complete probability space. $R$ is the random interval-valued return of a risky asset. Given confidence level $1-\alpha \  (0<\alpha<1)$, $IVaR$ is the Value-at-Risk. The conditional Value-at-Risk is defined by
	$$
	ICVaR:=-E(R|R
	\prec-IVaR).
	$$
\end{definition}
Similar to the real-valued CVaR, ICVaR describes the tail risk.
\begin{remark}\label{coherent}
	The expectation of random interval is linear  for positive coefficients (c.f.\cite{LiOgV}). As a manner similar to the real-valued CVaR, it can be proved that the risk measure ICVaR satisfies the sub-additivity.
\end{remark}

\subsection{Portfolio Models}

Let the rate of return of asset $R=[r^{L}, r^{U}]$ be an interval-valued random variable,
$R_{ij}=[r^{L}_{ij}, r^{U}_{ij}]$ be the rate of return of the $i$-th security at the $j$-th period, $i=1,\cdots, n$, $j=1,\cdots,k.$

Given the confidence level $1-\alpha$, $IVaR_{ij}$ ($ICVaR_{ij}$) is the  Value-at-Risk (conditional Value-at-Risk, resp.)  of the $i$-th risky asset at the $j$-th period, $i=1,\cdots, n$, $j=1,\cdots,k$. Let $x_{i}$ ($ i=1,\cdots,10$) be the proportion of the $i$-th asset in the portfolio. By Proposition \ref{coherent}, the $ICVaR$ of portfolio is less than the linear composition of $ICVaR_{i}$ with weights $x_{i}\geq 0$ ($i=1,\cdots,n$).

Now we build two models of portfolio selection as follows:
\begin{itemize}
	\item Model 1
	
	We consider the optimal portfolio problem under given acceptable  maximum of risk level:
	\begin{equation}\label{maxreturn}
		\begin{array}{ll}
			Maximize_{\circledless} \ \ & E(R_{P})={\sum\limits_{i=1}^n}x_{i}E(R_{i}),\\
			subject \ to \ \ &
			{\sum\limits_{i=1}^n}[ICVaR^{L}_{ij},ICVaR^{U}_{ij}]
			x_{i}\leq [ICVaR^{L}_{0j}, ICVaR^{U}_{0j}],\\
			\ \              &  j=1,\cdots, k,\\
			\ \              &  \sum\limits_{i=1}^{n}x_{i}=1, \ \ \ x_{i} \geq 0, \ i=1,\cdots,n.
		\end{array}
	\end{equation}
	where $E(R_{i})$ is the expectation return of the $i$-th asset, $ICVaR_{0j}=[ICVaR_{0j}^{L}, ICVaR_{0j}^{U}]$ is the acceptable maximum risk in the $j$-period, $j=1,\cdots, k$, which are subjective and given by the investors.
	
	By \eqref{maxreal}, the corresponding real-valued linear programming problem of \eqref{maxreturn} is represented as \eqref{maxreturnreal} below
	\begin{eqnarray}\label{maxreturnreal}
		\begin{array}{ll}
			Maximize \ \ & m(E(R_{P}))=
			{\sum\limits_{i=1}^n}x_{i}m(E(R_{i})),\\
			subject \ to \ \ & {\sum\limits_{i=1}^n} x_{i}ICVaR^{U}_{ij}
			\leq  ICVaR^{U}_{0j}, \\
			\ \  \ \ &
			{\sum\limits_{i=1}^n}x_{i}[m(ICVaR_{ij})-\gamma w(ICVaR_{ij})] \leq m(ICVaR_{0j}) + \gamma w(ICVaR_{0j}) ,\\
			\ & \ j=1,\cdots,k\\
			\ \ \ \ &  x_{i} \geq 0,\ for \ i=1,\cdots,n,
		\end{array}
	\end{eqnarray}
	where $m(A)$ is the midpoint of interval $A$ and $w(A)$ the semi-width of $A$. $\gamma \in(0,1)$ is a given index in advance, which can describe
	the degree of risk appetite of investors. The larger $\gamma $, the lower the risk aversion.
	
	\item
	Model 2
	\begin{equation}\label{minrisk}
		\begin{array}{ll}
			Minimize_{\circledless} \ \ & {\sum\limits_{i=1}^n}x_{i}ICVaR_{i},\\
			subject \ to \ \ &
			{\sum\limits_{i=1}^n}
			x_{i}E(R_{ij})\geq R_{0j},\\
			\ \              &  j=1,\cdots, k\\
			\ \              &  \sum\limits_{i=1}^{n}x_{i}=1, \ \ \ x_{i} \geq 0, \ i=1,\cdots,n.
		\end{array}
	\end{equation}
	
	Correspondingly, by \eqref{minreal}, the interval-valued linear programming problem \eqref{minrisk} is represented as the following real-valued linear programming model \eqref{minriskreal} below
	\begin{eqnarray}\label{minriskreal}
		\begin{array}{ll}
			Minimize \ \ & \frac{1}{2}
			{\sum\limits_{i=1}^n}x_{i}(ICVaR^{L}_{i}+ICVaR^{U}_{i}),\\
			subject \ to \ \ & {\sum\limits_{i=1}^n} x_{i}E(r^{L}_{ij})
			\geq  r^{L}_{0j}, \\
			\ \  \ \ &
			{\sum\limits_{i=1}^n}x_{i}[m(E(R_{ij}))+\gamma w(E(R_{ij}))]\geq m(R_{0j}) - \gamma w(R_{0j}) ,\\
			\ & \ j=1,\cdots,k\\
			\ \ \ \ &  x_{i} \geq 0,\ for \ i=1,\cdots,n,
		\end{array}
	\end{eqnarray}
	where $m, w$ and $\gamma$ have the same meaning as that in \eqref{maxreturnreal}.
	
\end{itemize}
\begin{remark} Due to the complexity of practical market and random intervals, it is difficult to get the distribution function of the return of a risk asset. Therefore it is almost impossible to calculate the exact values of $E(R_{ij})$,  $E(R_{i})$, $IVaR_{i}$, $ICVaR_{i}$ etc. In the following case study, we use historical data to estimate the related values. For mathematical expectation, the moment estimator \eqref{momentestimator} is employed, which is consistent. I.e. let $r_{t}, t=1,\cdots, T$ be the given observations of random variable $R$, the estimation of $E(R)$ is given as follows
	\begin{equation}\label{momentestimator}
		\hat{E(R)}=\frac{1}{T}\Sigma_{t=1}^{T}r_{t}.
	\end{equation}
	
\end{remark}

\section{Case Study}

This paper selects the stocks of ten listed companies of financial markets in Chinese mainland as samples for a case study. The listed companies  include oil, bank, coal, liquor,  security, medicine, technology industry and so on.
The ticker symbols of stocks  are listed in Table \ref{asset list}.  Daily prices including  the daily closing price $S_{j}$, the highest price $S_{j}^{U}$ and the lowest price $S_{j}^{L}$, which are collected from Jan. 2016 to Sept. 2020 (Data source: http://www.finance.sina.cn). The software R is employed to make the empirical analysis.
\begin{table}[!htbp]
	\caption{Sample Assets}
	\centering
	\begin{tabular}[h]{|c|ccccc|c|}
		\hline Number &ST01 &ST02 &ST03 &ST04 &ST05  \\
		\hline Ticker &SH600028 &SH600085 &SH600188 &SH600536 &SH601939\\
		\hline Number &ST06 &ST07 &ST08 &ST09 &ST10  \\
		\hline Ticker &SZ000333 &SZ6000735 &SZ000776 &SZ000858 &SZ002428\\
		\hline
	\end{tabular}
	\label{asset list}
\end{table}

\begin{table}[!htbp]
	\centering
	\caption{Jarque-Bera test}
	\setlength{\tabcolsep}{5mm}
	\begin{tabular}[h]{|c|ccccc|}
		\hline Number &ST01&ST02&ST03&ST04&ST05\\
		\hline  $\chi^2$ &$843.55$ &$1372.1$ &$611.96$ &$123.05$ &$1285.1$  \\
		\hline Number &ST06 &ST07 &ST08 &ST09 &ST10\\
		\hline  $\chi^2$ &$313.1$ &$292.48$ &$1883.1$ &$205.07$ &$138.27$  \\
		\hline  
		\multicolumn{1}{|c|}{} & \multicolumn{5}{c|}{$df=2,p<2.2\times 10^{-16}$}\\
		\hline
	\end{tabular}
	\label{JBtest}
\end{table}

According to the daily closing price, the highest price and the lowest price
of the stock, we calculate the exact value and the interval value of return rate. Let the real valued return $r_{j}=\ln S_{j}-\ln S_{j-1}$, $r^{L}_{j}=\ln S_{j}^{L}-\ln S_{j-1}$, $r^{U}_{j}=\ln S_{j}^{U}-\ln S_{j-1}$, $R_{j}=[r^{L}_{j}, r_{j}^{U}]$ be the interval return. As usual, the real valued return is not normal distributed, see  Table \ref{JBtest}. Then it is not suitable to compute the real valued VaR by using the variance-covariance method. The
historical simulation method is employed to estimate the real-valued VaR, CVaR and  IVaR, ICVaR. There are many methods to rank intervals. Here  the ``mean-first, left-second'' of intervals ranking
method is used to estimate IVaR and ICVaR under the confidence level $1-\alpha=0.95$. By the satisfactory crisp equivalent system, interval-valued programming problems are transformed into real-valued ones. Then the optimal solutions of the corresponding portfolio models are
obtained.

\begin{table}[!htbp]
	\caption{Portfolio, Model 1 with the same $ICVaR_{0j}, j=1,\cdots, 5$}
	\label{model 1 with same bound}
	\resizebox{\textwidth}{!}{
		\begin{tabular}{ccccccccccccc}
			\hline
			\multicolumn{2}{c}{$ICVaR_{0j}=[0.008,0.08], j=1,\cdots, 5$}\\
			
			\hline
			&ST01&ST02&ST03&ST04&ST05&ST06&ST07&ST08&ST09&ST10&$ST_{OP}$& $ \gamma$\\
			\hline
			[1]&0&0&0&0&0&0&0&0&1&0&0.0009 &0.15\\
			
			[2]&0&0&0&0&0&0.0349 &0&0&0.9651 &0&0.0009 &0.05\\
			
			[3]&0&0&0&0&0&0.0909 &0&0&0.9091 &0&0.0009 &0.04\\
			
			[4]&0&0&0&0&0&0.1462 &0&0&0.8538 &0&0.0009 &0.03\\
			
			[5]&0&0&0&0&0&0.1736 &0&0&0.8264 &0&0.0009 &0.025\\
			
			[6]&0&0&0&0&0&0.2008 &0&0&0.7992 &0&0.0009 &0.02\\
			
			[7]&0&0&0&0&0&0.2869 &0&0&0.7131 &0&0.0009 &0.01\\
			\hline
			
			\multicolumn{2}{c}{$ICVaR_{0j}=[0.003,0.07], j=1,\cdots, 5$}\\
			
			\hline
			&ST01&ST02&ST03&ST04&ST05&ST06&ST07&ST08&ST09&ST10&$ST_{OP}$&$\gamma$\\
			\hline
			[1]&0&0&0&0&0.1537 &0.7908 &0&0.0000 &0.0554 &0&0.0007 &0.15\\
			
			[2]&0&0&0&0&0.1875 &0.1268 &0&0.3520 &0.3337 &0&0.0005 &0.05\\
			
			[3]&0&0&0&0&0.2136 &0.0973 &0&0.3935 &0.2956 &0&0.0004 &0.04\\
			
			[4]&0&0&0&0&0.2395 &0.0680 &0&0.4348 &0.2577 &0&0.0004 &0.03\\
			
			[5]&0&0&0&0&0.2523 &0.0535 &0&0.4552 &0.2389 &0&0.0004 &0.025\\
			
			[6]&0&0&0&0&0.2652 &0.0389 &0&0.4755 &0.2203 &0&0.0003 &0.02\\
			
			[7]&0&0&0&0&0.3270 &0&0&0.4792 &0.1938 &0&0.0002 &0.01\\
			\hline
		\end{tabular}
	}
\end{table}

In Table \ref{model 1 with same bound}, when the maximum risk level $ICVaR_{0}$ is different, the selection of assets is different. When taking the lower acceptable risk level $ICVaR_{j0}=[0.003, 0.07]$, the investor prefers to the stock ST05 (CCB, China Construction Bank) and ST08 (GF Securities Co., Ltd) since these two stocks have stable returns and lower risk compared with the stock ST06 and ST09, which implies the investor is more conservative than one who take $ICVaR_{0j}=[0.008, 0.08]$. Table \ref{model 1 with same bound} also shows that the less $\gamma$, the more allocation to assets with lower risk. Correspondingly, the weights of ST06 (Mediea Group) and ST09 (Wuliangye Group) increase with respect to $\gamma$. The result implies that ST06 and ST09 have higher return but accompanying with higher risk, which is consistent with the actual situation of the market.

\begin{table}[!htbp]
	\caption{Portfolio, Model 2 with the same $R_{0j}, j=1,\cdots, 5$}
	\label{model 2 with same bound}
	\resizebox{\textwidth}{!}{
		\begin{tabular}{ccccccccccccc}
			\hline
			\multicolumn{2}{c}{$R_{0j}=[-0.025,0.025], j=1,\cdots, 5$}\\
			
			\hline
			&ST01&ST02&ST03&ST04&ST05&ST06&ST07&ST08&ST09&ST10&$ST_{OP}$ &$\gamma$\\
			\hline
			{[1]}&0&0&0&0&1&0&0&0&0&0&0.0243 &0.04\\  \relax
			[2]&0&0&0&0&0.9045 &0&0.0955 &0&0&0&0.0265 &0.03\\ \relax
			[3]&0&0&0&0&0.8332 &0&0.1668 &0&0&0&0.0281 &0.025\\ \relax
			[4]&0&0&0&0&0.7253 &0.0285 &0.2462 &0&0&0&0.0302 &0.02\\ \relax
			[5]&0&0&0&0&0.3844 &0.1797 &0.4358 &0&0&0&0.0361 &0.01\\
			\hline	
			\multicolumn{2}{c}{$R_{0j}=[-0.02,0.02], j=1,\cdots, 5$}\\
			
			\hline
			&ST01&ST02&ST03&ST04&ST05&ST06&ST07&ST08&ST09&ST10&ST\_OP&$\gamma$\\
			\hline
			{[1]}&0&0&0&0&0.9717 &0&0.0283 &0&0&0&0.0250 &0.04\\ \relax
			[2]&0&0&0&0&0.8534 &0&0.1466 &0&0&0&0.0277 &0.03\\ \relax
			[3]&0&0&0&0&0.7825 &0.0062 &0.2113 &0&0&0&0.0292 &0.025\\ \relax
			[4]&0&0&0&0&0.6458 &0.0665 &0.2877 &0&0&0&0.0316 &0.02\\
			\hline
	\end{tabular}}
\end{table}
\begin{table}[!htbp]
	\caption{Portfolio, Model 1 with different $ICVaR_{0j}$ for different $j=1, \cdots, 5$}
	\label{model 1 with different bound}
	\resizebox{\textwidth}{!}{
		\begin{tabular}{ccccccccccccc}
			\hline
			\multicolumn{3}{c}{$ICVaR_{01}=[0.004,0.08]$}&\multicolumn{3}{c}{$ ICVaR_{02}=[0.0045,0.06]$}&
			\multicolumn{3}{c}{$ICVaR_{03}=[0.007,0.08]$}&\multicolumn{4}{c}{}\\
			
			\multicolumn{3}{c}{$ICVaR_{04}=[0.007,0.06]$}&\multicolumn{3}{c}{$ICVaR_{05}=[0.018,0.07]$}&\multicolumn{7}{c}{}\\
			
			\hline
			& ST01&ST02&ST03&ST04&ST05&ST06&ST07&ST08&ST09&ST10&$ST_{OP}$&$\gamma$\\
			\hline
			[1]&0&0&0&0&0&0.1700 &0&0&0.8300 &0&0.0009 &0.05\\
			
			[2]&0&0&0&0&0&0.1700 &0&0&0.8300 &0&0.0009 &0.04\\
			
			[3]&0&0&0&0&0&0.1734 &0&0&0.8266 &0&0.0009 &0.03\\
			
			[4]&0&0&0&0&0&0.1962 &0&0&0.8038 &0&0.0009 &0.025\\
			
			[5]&0&0&0&0&0&0.2189 &0&0&0.7811 &0&0.0009 &0.02\\
			
			[6]&0&0&0&0&0&0.8222 &0&0&0.1778 &0&0.0009 &0.01\\
			\hline
		\end{tabular}
	}
\end{table}

\begin{table}[!htbp]
	\caption{Portfolio, Model 2 with different $R_{0j}$ for different $j=1,\cdots, 5$}
	\label{model 2 with different bound}
	\resizebox{\textwidth}{!}{
		\begin{tabular}{ccccccccccccc}
			\hline
			
			\multicolumn{3}{c}{$R_{01}=[-0.020,0.025]$}&\multicolumn{3}{c}{$R_{02}=[-0.020,0.020]$}&
			\multicolumn{3}{c}{$R_{03}=[-0.025,0.025]$}&\multicolumn{4}{c}{}\\
			
			\multicolumn{3}{c}{$R_{04}=[-0.025,0.020]$}&\multicolumn{3}{c}{$R_{05}=[-0.021,0.016]$}&\multicolumn{7}{c}{}\\
			
			\hline
			&ST01&ST02&ST03&ST04&ST05&ST06&ST07&ST08&ST09&ST10&$ST_{OP}$&$\gamma$\\
			\hline
			[1]&0&0&0&0&1&0&0&0&0&0&0.0243 &0.15\\
			
			[2]&0&0&0&0&0.5045 &0.4955 &0&0&0&0&0.0295 &0.05\\
			
			[3]&0&0&0&0&0.3512 &0.6155 &0.0333 &0&0&0&0.0315 &0.04\\
			
			[4]&0&0&0&0&0.0533 &0.7534 &0.1933 &0&0&0&0.0366 &0.03\\
			\hline
		\end{tabular}
	}
\end{table}

In Table \ref{model 2 with same bound}, given the minimum return constraint $R_{0j}=[-0.025, 0.025]$ and $[-0.02, 0.02]$, to minimize the risk, it is obvious that ST05 has the highest proportion in each scenario, very different from the case in Model 1.  Its lower risk property explains this result. Also one can figure out that the weight of stock ST05 increases with respect to $\gamma$ however the proportion of ST07 behaves in the contrary. In a word, the solution is sensitive to the constraints and also depends on the index $\gamma$. One also can figure out the law from Table \ref{model 1 with different bound} and Table \ref{model 2 with different bound}.

Different risk preference has different weights for portfolio selection.
In the process of solving the model, we can reasonably allocate the investment share
according to the degree of risk preference of investors. So that it is important to select appropriate constraints and $\gamma$ for making a decision of portfolio selection for the investors.
\section{Concluding remarks}
This paper contributes to the optimal selection under interval-valued environment.  At first, the interval-valued conditional Value-at-Risk (ICVaR) was defined, which satisfies the sub-additivity. An then, based on the  risk measure, two interval-valued portfolio selection models were constructed, where both the return and risk of a risky asset are random intervals. The``mean-first, left-second" method was employed to rank the intervals for estimating IVaR and ICVaR by historical simulation method.  Of course, one can use other methods to rank intervals, such as "mean-first, right-second" or possibility degree etc. Different ranking method will lead to different IVaR and ICVaR for the risky asset. On solving the maximization and minimization problem of interval values, the  satisfactory crisp equivalent system of $Ax\leq B$ (\cite{Sen}) was used. In Model 1 and Model 2, the $\Gamma$-index can describe  investors' subjective preference or aversion to risk, which is an innovation different from the classical portfolio model. The smaller the index, the higher the risk aversion of investors. The result of case study also shows it. But the real finance market is very complicated. The risk of asset is not only
related to the industry, but also closely related to the size of the company, trading time, national
policy, and so on. It is expected that the models have certain guiding significance for investors' choice.


\section*{Acknowledgment}
  The research of Jinping Zhang was supported by Natural Science Foundation of Beijing Municipality (No.1192015).

\bibliographystyle{plain}

\end{document}